\documentclass[aps,secnumarabic,eqsecnum,showpacs,nobibnotes,amsmath,
eqsecnum,amssymb,prd]{revtex4}
\newcommand{\be}{\begin{equation}}
\newcommand{\ee}{\end{equation}}
\newcommand{\bq}{\begin{eqnarray}}
\newcommand{\eq}{\begin{eqnarray}}
\begin{document}
\title{\large\bf Three Generations of SUSY Standard 
Model of Nambu-Goto  String}
\medskip
\author{B. B. Deo }
\affiliation{ Physics Department, Utkal University, 
Bhubaneswar-751004, India.}
\begin{abstract}
A four dimensional Superstring have been constructed 
starting from a twenty six dimensional bosonic string.
Fermions are introduced by noting the Mandelstam's 
proof of equivalence of 
two fermions to one boson in 1+1 dimensions. The 
action of the superstring 
is invariant under $ SO(6) \otimes SO(5)$ in
the world sheet.It has four bosonic coordinates and forty four  
Majorana fermions of $SO(3,1)$. 
Here the superstring action is shown to be invariant 
under conformal and superconformal transformations with the usual
conformal and superconformal ghosts of the 10-d superstrings. 
The massless spectrum obtained by
quantising the action, contain vector mesons which are generators 
of the $SO(6) \otimes SO(5)$.
Using Wilson loops, this product group is proven to descend to 
$Z_3\otimes SU(3)\otimes SU(2)\otimes U(1)$ without 
breaking supersymmetry. Thus there are just three generations of 
quarks and leptons.
\end{abstract}
\pacs{11.25-w, 11.30Pb, 12.60.Jv}
\date{\today}
\maketitle
\section{Introduction}
Recently, we have shown that a novel four dimensional Superstring 
can be obtained from the basic Nambu-Goto
bosonic string. In this paper, we report the construction of the 
same $N=1$ four dimensional superstring in which the spectrum of 
excitations includes  three generations of the Standard Model. 
The theory is anomaly free, modular invariant, free of ghosts 
and is unitary. Some salient features
of the model are outlined below.

In the earlier work on this topic as contained in 
references ~\cite{Deo03} and ~\cite{Deo103},
the importance of superconformal invariance for 
the superstring had not been stressed
for arbitrary genus g$>$0. We retrieve the 
superconformal ghosts and show how they cancel the anomaly. 
The normal quantum ordering constant remain the same
$a$=-1, eventhough there are two equivalent ways of writing down
the total super Virasoro energy momentum generator. This is 
given in Section-4. There have been many attempts to study 
four dimensional strings ~\cite{Dine88}, specially from the 
later half of the eighties. 
Antoniadis et al~\cite{Antoniadis87,Chang88,Kawai87}
have used eighteen fermions with four bosons in trilinear coupling 
and Chang and Kumar with Thirring
interaction. The situation and methodology, as prevailed in 1988/1989, 
has been given in reference~
\cite{Lerche89} with many specific examples of construction.

Not long ago, Casher, Englert, Nicolai and Taormina ~\cite{Casher85} 
showed that consistent superstring can be solutions of 
$D=26$ bosonic strings and the latter appears to 
be the fundamental string theory. This has 
been persued further in~\cite{Casher85} and~ 
\cite{Englert86, Freund85,Chattaraputi02,Chattaraputi021}.
The latest approaches have been to derive the conventional 
ten dimensional superstring from 
the 26-d bosonic string and then compactify to reach out 
to phenomenology. There are several $U(1)$s.
The present formulation of the theory is quite different 
in calculational details and results, from those given in 
earlier references importantly from ~\cite{Antoniadis87} 
to \cite{Casher85}. For completeness, we review some major
steps given in our earlier references.

The model, which we have proposed, begins from the 
Nambu-Goto~\cite{Nambu70, Goto71} bosonic string theory in 
the world-sheet ($\sigma ,\tau$) in 26 dimensions. 
The reason for the dimensionality is easy to see. The string 
action in twenty six dimension is
\begin{equation}
S_B=-\frac{1}{2\pi}\int d^2\sigma 
\left ( \partial_{\alpha}X^{\mu}\partial^{\alpha} X_{\mu} \right ),~~~\mu 
=0,1,2,...,25,\label{e1}
\end{equation}
where $\partial_{\alpha}=(\partial_{\sigma}, \partial_{\tau})$. 
The central charge for bosons is found by using the general 
expression for the two energy momentum tensors at two world 
sheet points z and $\omega$

\begin{equation}
2< T(z)T(\omega) > = \frac{C}{(z-\omega)^4} + ...\label{e2}
\end{equation}
The coefficient of the most divergent term as $C$ in equation (\ref{e2}) 
is the central charge. Methods and principles of calculation of the 
central charge and those for a variety of strings 
has been given in reference~\cite{Lerche89}. For free bosons, the central 
charge is

\begin{equation}
C_B=\delta_{\mu}^{\mu}.\label{e3}
\end{equation}

The action $S_B$ in equation (\ref{e1})  has a central charge 
$\delta_{\mu}^{\mu}$=26 and  hence it  is not anomaly free. 
One adds to equation (\ref{e1}) the action of conformal 
ghosts $(c^+, b_{++})$ and the generator with quanta $(c,b)$,

\begin{equation} 
S_{FP}=\frac{1}{\pi} \int
\left ( c^+\partial_-b_{++}+ c^-\partial_+b_{--}\right ) d^2\sigma ,
~~~~~~~~ L_{m}^{FP}= \sum_n(m-n)b_{m+n}c_{-n}.\label{e4}
\end{equation}

This action has  a central charge $-26$, independent of the 
dimensionality of the string. To have an
anomaly free string theory, the central charge of the 
conformal ghosts should be able to cancel 
and this happens when the $C=\delta^{\mu}_{\mu} =D=26$. So the 
string is physical only in $D=26$ dimensions with the total 
central charge zero.
Using Mandelstam's~\cite{Mandelstam75} proof of equivalence 
between one boson to two fermionic 
modes in the infinite volume limit in 1+1 dimensional 
field theory, one can rewrite the action 
as the sum of the four bosonic coordinates $X^{\mu}$ of $SO(3,1)$ 
and forty four fermions having internal symmetry $SO(44)$, 
thereby loosing Lorentz invariance.
If this is anomaly free, this is also true 
in finite intervals or a circle.
Noting that the Majorana fermions can be in bosonic representation 
of the Lorentz group $SO(3,1)$, 
the forty four fermions are grouped into eleven Lorentz vectors 
of $SO(3,1)$ which look as a commuting
internal symmetry group when viewed from the other internal 
quantum number space. The action is now
\begin{equation}
S_{FB}=-\frac{1}{2\pi}~\int~d^2 \sigma 
\left [ \partial^{\alpha}X^{\mu}(\sigma,\tau)~
\partial_{\alpha}X_{\mu}(\sigma,\tau) - 
i\sum_{j=1}^{11}\bar{\psi}^{\mu,j} 
\rho^{\alpha}\partial _{\alpha}\psi_{\mu,j} \right ],\label{e5}
\end{equation}
and is anomaly free with $S_{FP}$ of equation (\ref{e4}). The 
upper indices $j,k$ refer to a row and that of lower to a column, and
\begin{eqnarray}
\rho^0 =
\left (
\begin{array}{cc}
0 & -i\\
i & 0\\
\end{array}
\right )
\end{eqnarray}
and
\begin{eqnarray}
\rho^1 =
\left (
\begin{array}{cc}
0 & i\\
i & 0\\
\end{array}
\right ).
\end{eqnarray}
Dropping indices
\begin{equation}
\bar{\psi}=\psi^{\dag}\rho^0.
\end{equation}

Here $\rho^{\alpha}$'s are imaginary, so the Dirac operators 
$\rho^{\alpha}\partial_{\alpha}$ are real. In 
this representation of Dirac algebra, the components of the world 
sheet spinor $\psi^{\mu,j}$ are real and they are Majorana spinors.

One has introduced an anticommuting field $\psi^{\mu,j}$ that transforms 
as vectors - a bosonic 
representation of $SO(3,1)$. $\psi^{\mu}_A$ maps bosons to bosons 
and fermions to fermions in the 
space-time sense. There is no clash with spin statistics theorem. 
Action (\ref{e5}) is a two dimensional
field theory, not a  field theory in space time. $\psi^{\mu}_A$ transforms 
as a spinor under the 
transformation of the two dimensional world sheet. The Lorentz group 
$SO(3,1)$ is merely an internal symmetry
group as viewed in the world sheet as stated earlier. This is discussed 
in reference \cite{Green87}.

The central charge of the  free fermions in action (\ref{e5}) as 
deduced by calculation from equation
(\ref{e2}) is
\begin{equation}
C_F= \frac{1}{2} \delta^{\mu}_{\mu}\delta_j^j\label{e1a},
\end{equation}
so that the total central charge of equation (\ref{e5}) is
\begin{equation}
C_{ SB}=\delta_{\mu}^{\mu} + \delta_{\mu}^{\mu}\delta_j^j 
= 4+ ( \frac{1}{2}\times 4\times 11)=26\label{e6}
\end{equation}
It appears that the conformal ghosts will cancel this. 
As a check of the equation (\ref{e6}), 
the central charge of a ten dimensional 
superstring is 10+($\frac{1}{2}\times 10\times 1)=15$ . 
This is correct.

\section{Supersymmetry}

The action (\ref{e5}), however, is not supersymmetric. 
The eleven $\psi^{\mu,j}_A$ have to be further 
divided into two species; $\psi^{\mu,j} ,~~ j=1,2,...,6$ 
and $\phi^{\mu, k},~~k=7,8,...,11$. 
For the group of six, the positive and negative parts of 
$\psi^{\mu,j}=\psi^{(+)\mu,j} +\psi^{(-)\mu,j}$,  
whereas for the group of five, allowed the freedom of 
phase of creation operators for Majorana fermions
in $\phi^{\mu,k} = \phi^{(+)\mu,k}-\phi^{(-)\mu,k}$. The action is now

\begin{equation}
S= -\frac{1}{2\pi} \int d^2\sigma
\left [\partial_{\alpha}X^{\mu}\partial^{\alpha}X_{\mu}
-i~\bar{\psi}^{\mu,j}~\rho^{\alpha}~\partial_{\alpha}~\psi_{\mu,j}
+ i~\bar{\phi}^{\mu,k}~\rho^{\alpha}~
\partial_{\alpha}~\phi_{\mu,k}\right ].
\label{e7}
\end{equation}

Besides $SO(3,1)$, the action (\ref{e7}) is invariant 
under $SO(6)\otimes SO(5)$. It is also 
invariant under the supersymmetric transformation
\begin{eqnarray}
\delta X^{\mu} =\bar{\epsilon}~(e^j\psi^{\mu}_j - e^k\phi^{\mu}_k),\\
\delta\psi^{\mu,j}
= - ie^j\rho^{\alpha}\partial_{\alpha}X^{\mu}~\epsilon,\\
and~~~~~\delta\phi^{\mu,k}
= ie^k\rho^{\alpha}\partial_{\alpha}X^{\mu}~\epsilon.\label{e8}
\end{eqnarray}
Here $\epsilon$ is a constant anticommuting spinor. 
$e^j$ and $e^k$ are eleven numbers of a 
row with $e^j e_j$=6 and $e^k e_k$=5. They look 
like~ $e^1$=(1,0,0,0,0,0,0,0,0,0,0) ~and
~$e^7$=(0,0,0,0,0,0,-1,0,0,0,0). Such arrays have been introduced 
in reference~\cite{Lerche89}
to obtain different group structure. Ours is a specific 
case with phenomenological applications.

There is wide mismatch between the fermionic and bosonic 
modes in the action (\ref{e7}).
To investgate this disturbing feature, we find that 
the commutators of two successive 
supersymmetric transformations 
lead to a translation with the coefficients 
$a^{\alpha} = 2~i~\bar{\epsilon}^1
\rho^{\alpha}\epsilon_2$ provided the internal symmetry 
indices in equation (\ref{e7}) satisfy
\begin{equation}
\psi_j^{\mu} = e_{j}\Psi^{\mu},~~~~~~~and~~~~~~
\phi_k^{\mu} = e_{ k}\Psi^{\mu}.\label{e9}
\end{equation}
These are the two key equations. These equations state 
that, of the eleven sites $(j,k)$, 
the super fermionic partner $\Psi^{\mu}$ is found in one site only.
On  j$^{th}$ or k$^{th}$ site, it emits or 
absorbs quanta as found by quantising $\psi^{\mu,j}$ or
$\phi^{\mu,k}$ respectively prescribed by the action (\ref{e7}). 
The alternative auxiliary 
fields are not needed. The $\Psi^{\mu}$ is given by the linear sum,
\begin{equation}
\Psi^{\mu} = e^j\psi^{\mu}_j - e^k\phi^{\mu}_k.\label{e10}
\end{equation}
It is easy to verify that
\begin{eqnarray}
\delta X^{\mu}=\bar{\epsilon}\Psi^{\mu}, \;\;\;\;\;\;\;\;\; 
\delta \Psi^{\mu}=-i\;
\epsilon\;\rho^{\alpha}\;\partial_{\alpha}\;X^{\mu}\label{e11}
\end{eqnarray}
and 
\begin{equation}
[\delta_1 ,\delta_2]X^{\mu } 
= a^{\alpha}\partial_{\alpha}X^{\mu },\;\;\;\;\;\;
[\delta_1 ,\delta_2]\Psi^{\mu } 
= a^{\alpha}\partial_{\alpha}\Psi^{\mu }.\label{e12}
\end{equation}
We immediately obtain a Nambu-Goto superstring in four 
dimensions from the action (\ref{e7})
using equation (\ref{e9}). The action is
\begin{equation}
S=-\frac{1}{2\pi}~\int~d^2\sigma
\left ( \partial_{\alpha} X^{\mu}\partial^{\alpha}X_{\mu}
-~i\Psi^{\mu}\rho^{\alpha}\partial_{\alpha}\Psi_{\mu}\right )\label{e13}
\end{equation}

Thus the action(\ref{e7}) is truely supersymmetric and 
is also the equivalent action of a superstring 
in four dimension. Quantising this action in well written 
down procedure, the particle spectrum is very rich unlike 
quantising (\ref{e13}). We also need the internal symmetry 
$SO(6)\otimes SO(5)$ of the action (\ref{e7}).

\section{superconformal  invariance}

In the formulation of superstring theory in the above section, 
a principal role has been
played by the proof that the commutator of two supersymmetric 
transformations gives a world
sheet translation. Therefore, it is necessary to have an 
exact framework in which the super
Virasoro conditions can emerge as gauge conditions. For this, 
the action given in (\ref{e7}) 
should incorporate the superconformal invariance of a full 
superstring theory.

From equations (\ref{e11}) and (\ref{e12}), it follows 
that the superpartner of $X^{\mu}$
is $\Psi^{\mu}$. Introducing another supersymmetric pair, the Zweibein 
$e^{\alpha}(\sigma,\tau)$ and the gravitons 
$\chi_{\alpha}= \nabla_{\alpha}\epsilon$,
the local 2-d supersymmetric action, first written down by 
Brink, Di Vecchia, Howe, Deser
and Zumino~\cite{Brink76, Deser76} is
\begin{equation}
S= -\frac{1}{2\pi}\int d^2\sigma ~e~
\left [ h^{\alpha\beta}\partial_{\alpha}X^{\mu }
\partial_{\beta}X_{\mu } -i\bar \Psi^{\mu}
\rho^{\alpha}\partial_{\alpha}
\bar \Psi_{\mu}+ 2\bar{\chi}_{\alpha}\rho^{\beta}
\rho^{\alpha}\Psi^{\mu}
\partial_{\beta}\chi^{\mu}+\frac{1}{2}
\bar{\Psi }^{\mu}\Psi_{\mu}\bar{\chi}_{\beta} 
\rho^{\beta}\rho^{\alpha}\chi_{\alpha}
\right ].\label{e14}
\end{equation}
A detailed derivation is given in reference \cite{Green87}. 
The Einstein-Hilbert action
$\int {eR~~d^2\sigma}$ can be added, but this does not 
change the classical analysis. The action is invariant
under local supersymmetric transformations
 \begin{equation}
 \delta X^\mu=\bar\epsilon \Psi^\mu,~~~~~~~~and~~~~~~~~~
 \delta \Psi^\mu=-i\rho^{\alpha} \epsilon
\left (\partial_{\alpha}X^{\mu}-\bar{\Psi}^{\mu}
 \chi_{\alpha}\right),
 \end{equation}
and
 \begin{equation}
 \delta e^{a}_{\alpha}=-2i\bar\epsilon \rho^{a}\chi_{\alpha},~~~~~~~~~
  \delta \chi_{\alpha}=\nabla_{\alpha}\epsilon.
 \end{equation}
 There are two other important transformations,\\
(a) the Weyl transformations
 \begin{equation}
 \delta X^{\mu}=0,~~~~~~~~~~~\delta \Psi^{\mu}=
- \frac{1}{2} \Lambda \Psi^{\mu},
 \end{equation}
 \begin{equation}
 \delta e^{a}_{\alpha}=\Delta e^{a}_{\alpha},
~~~~and~~~~~~\delta \chi_{\alpha}
 =\frac{1}{2}\Lambda \chi _{\alpha},
 \end{equation}
 and \\
(b) the local fermionic symmetry, with  $\eta$ , an arbitrary 
Majorana spinor
\begin{equation}
\delta \chi _{\alpha}=i\rho_{\alpha} \eta,\label{e25a}
\end{equation}
and
\begin{equation}
\delta e^{a}_{\alpha}=\delta \psi^{\mu}=\delta X^{\mu}=0.\label{e26a}
\end{equation}

The invariance (b) requires the identity 
$\rho^{\alpha} \rho_{\beta}\rho_{\alpha}=0$,
which is true in two dimensions. All these
transformation properties imply that the 
action(\ref{e14}) is superconformal invariant.
Varying the field and Zweibein, the Noether current $J^{\alpha}$ and 
the energy momentum tensor $T_{\alpha\beta}$ vanishes,
\begin{equation}
J_{\alpha}= \frac{\pi}{2e}\frac{\delta S}{\delta \chi^{\alpha}}
=\rho^{\beta}
\rho_{\alpha}\bar{\Psi}^{\mu}
\partial_{\beta}X_{\mu}=0,\label{e14a}
\end{equation}
and
\begin{equation}
T_{\alpha\beta}=\partial_{\alpha}X^{\mu }
\partial_{\beta}X_{\mu }- 
\frac{i}{2}\bar{\Psi}^{\mu}\rho_{(\alpha}\partial_{\beta )}
\Psi_{\mu}=0.\label{e15a}
\end{equation}
These are the super Virasoro constrain equations as derived 
from the algebra.

In a light cone basis, the vanishing of the lightcone
components are obtained from variation of the action(\ref{e14}) 
i.e. equations (\ref{e14a}) and (\ref{e15a})
\begin{equation}
J_{\pm}=\partial_{\pm}X_{\mu}\Psi^{\mu}_{\pm}=0,\label{e16a}
\end{equation}
and
\begin{equation}
T_{\pm\pm}=
\partial_{\pm}X^{\mu}\partial_{\pm}X_{\mu}
+\frac{i}{ 2}\psi^{\mu j}_{\pm}\partial_{\pm}
\psi_{\pm\mu,j }- 
\frac{i}{2}\phi_{\pm}^{\mu k}\partial_{\pm}\phi_{\pm\mu,k},\label{e15}
\end{equation}
where $\partial_{\pm}=\frac{1}{2}(\partial_{\tau} \pm\partial_{\sigma})$. 

The action in equation (\ref{e14}) is not space-time 
supersymmetric. However, in the 
fermionic representation SO(3,1), fermions are Dirac 
spinor with four components $\alpha$. 
We construct Dirac spinor, like equation (\ref{e10}), 
as the sum of component spinors
\begin{equation}
\Theta_{\alpha}=\sum^6_{j=1}e^j\theta_{j\alpha} -
\sum^{11}_{k=7}e^k\theta_{k\alpha}.
\end{equation}
With the usual Dirac matrices $\Gamma^{\mu}$, since the identity
\begin{equation}
\Gamma_{\mu}\psi_{[1}\bar{\psi}_2\Gamma^{\mu}\psi_{3]} =0,
\end{equation}
is satisfied due to the Fierz transformation in four dimension, 
the Green-Schwarz action
~\cite{Green84} for $N=1$ supersymmetry is
\begin{equation}
S=\frac{1}{2\pi}\int d^2\sigma 
\left ( \sqrt{g}g^{\alpha\beta}\Pi_{\alpha}\Pi_{\beta}
+2i\epsilon^{\alpha\beta}\partial_{\alpha}X^{\mu}\bar{\Theta}\Gamma_{\mu}
\partial_{\beta}\Theta\right ),\label{e16}
\end{equation}
where 
\begin{equation}
\Pi^{\mu}_{\alpha}=\partial_{\alpha}X^{\mu}- 
i\bar{\Theta}\Gamma^{\mu}\partial_{\alpha}\Theta .\label{e17}
\end{equation}

This is the $N=1$ and $D=4$ superstring, originating from the
 $D=26$ bosonic string. It is 
difficult to quantise this action covariantly. It is 
better to use NS-R~\cite{Neveu71, Ramond71} 
formulation with G.S.O projection~\cite{Gliozzi76}. 
This has been done in references~\cite{Deo03, Deo103}.

\section{superconformal ghosts}
In a conformal ghost space equation(\ref{e4}), the path integration 
over metric $h_{\alpha \beta}$ could be
simply replaced by one over conformal factor $\phi$ and 
reparametrisation coordinates. This is 
true only if the given metric can be reached by such inputs. 
For world sheets of genus $g > 0$,
it is not possible to do so and reach all metrics from just one in 
this simple way. Something
more must be done using superconformal transformations.

The anticommuting ghost coordinates are present in this superstring. 
But the bosons are less
and fermions more. These are more intricate ones due to the 
local world sheet supersymmetry and superconformal invariance.

So we should attempt to isolate this superconformal ghost 
action following  well laid
down procedure for local world sheet supersymmetry. 
Actually, the gravitino $\chi_\alpha$ has
been gauged away. So this string is anomaly free, even without them. 
We shall put it back in the path integral and retrieve 
the superconformal ghost action. Using $\eta$  and $\epsilon$  
as in equations (\ref{e25a}) and (\ref{e26a}),
\begin{equation}
\chi_{\alpha}=i \rho_{\alpha} \eta + \nabla_{\alpha}\epsilon.
\end{equation}
The integration variable has been changed to that of $\eta$ and $\epsilon$, 
\begin{equation}
\delta \chi_{3/2}=\nabla_{1} \epsilon_{1/2},~~~~~~~~
~~~\delta \chi_{1/2}=\nabla_{1} \epsilon_{-1/2} + \eta_{1/2},
\end{equation}
and
\begin{equation}
\delta \chi_{-1/2}=\nabla_{-1} \epsilon_{1/2} + \eta_{-1/2},~~~~and~~~~~
\delta \chi_{-3/2}=\nabla_{-1} \epsilon_{-1/2}.
\end{equation}
Two of the above four are important. Changing variables 
from $\chi_{3/2}$ to $\epsilon_{1/2}$,
the Jacobian is ~\cite{Green87}
\begin{equation}
J_{3/2}=det^{-1} \nabla_{1}^{1/2 \rightarrow 3/2}
=\int D \gamma_{1/2}~~D \beta_{-3/2} exp \left( -\frac{1}{\pi} 
\int d^2 \sigma~~\beta_{-3/2}~\nabla_{1} \gamma_{1/2} \right),
\end{equation}
and from $\chi_{-3/2}$ to $\epsilon_{-1/2}$, the other Jacobian is,
\begin{equation}
J_{-3/2}=\int D \gamma_{-1/2} D \beta_{3/2}~~ exp \left(-\frac{1}{\pi}
\int d^2 \sigma~~\beta_{3/2}~\nabla_{-1} \gamma_{-1/2} \right).
\end{equation}
The resulting ghost action is~\cite{Green87}
\begin{equation}
S_{SC}=- \frac{1}{2 \pi} \int d^2 \sigma ~eh^{\alpha \beta} \bar {\gamma}
 \nabla_{\alpha} \beta_{\beta}.\label{e17a}
\end{equation}
The energy momentum tensor is
\begin{equation}
T_{++}=- \frac{1}{4} \gamma~ \partial_{+}\beta - 
\frac{3}{4} \beta~ \partial_{+} \gamma.
\end{equation}
This constitutes the central charge 11 ( or -11).
After the usual commutator quantisation, the superconformal 
ghost generator is
\begin{equation}
L_{m}^{gh,sc}= \sum \left( \frac{1}{2}m+n \right)~~ : 
\beta_{m-n} \gamma_{n}:~~.\label{e42a}
\end{equation}
Superconformal ghosts are necessary to build BRST physical states. 
With the appearance of these new ghosts, anomaly cancellation 
becomes subtler. The covariant
formulation result for the energy momentum correlation for one 
loop can be put in the form

\begin{equation}
2\left< T_{+}(\sigma)T_{+}(\sigma^{ \prime}) \right>
=\frac{1-3k^2}{(\sigma-\sigma^{ \prime})^4},
\end{equation}
so that the central charge for the one loop particle 
is $C=\pm (1-3k^2)$, depending on
its statistics. Here $k$ can be related to the conformal 
dimension $J=(1+k)/2$ . For each boson
$k=0$ and $C=1$, for conformal ghost $k=3$ and $C=-26$, and 
for the superconformal ghost $k=2$ 
and $C=11$ due to statistics. So the central 
charge of the four bosons of the action
(\ref{e7}) is 4, of the conformal ghost of equation 
(\ref{e4}) is $-26$ and of the
 superconformal ghost (\ref{e17a}) is 11. All these 
add up to $4-26+11=-11$. The 44
 fermions of the action (\ref{e7}) have been left out. 
To cancel this anomaly,  
each fermionic loop is to be characterised by 
$ k= \frac {1}{\sqrt{6}}$, so that the
central charge for each one of a pair is $1/4$ and 
hence the total contribution  
of the 44 fermionic mode is equal to 11. Thus there 
is a total cancellation of all
anomalies and, the superstring is anomaly free. 
Due to the gravitino consideration,
the action (\ref{e7}) of this novel superstring 
has central charge $4+11=15$, same as
that of the 10-d superstring, i.e., $10+5=15$.

Interestingly, this string theory falls into $N=1$ 
supersymmetry as per the table of the complete
list given by Polchinski~\cite{Polchinski98}, since $C^{gh}=-15$.
It is also appropriate to consider the gravitinos to have 
been gauged away and their 
contribution to the loop integral can be included due 
to the 44 Mandelstam Majorana
fermions. As they are normal fermions, $k=0$ or $J=1/2$, 
the 22 pairs contribute 22 to
the central charge. Together with 4 bosons, the central 
charge is 26, the same as that of the
Nambu-Goto string. We shall follow this alternative as 
they are near-identical descriptions.
The sum of all super Virasoro generators without any 
anomaly is expressed in the following equations.
\begin{eqnarray}
L_m^{sum}&=& L_m^{FP}+\frac{1}{\pi} 
\int _{- \pi}^{\pi} e^{im \sigma}~T_{++}~d\sigma\\
&=& L_m^{FP}+L_m^{gh,sc}+ \frac{1}{\pi} 
\int_{-\pi}^{pi} e ^{im \sigma}
\left( T_{++}-k \partial_{+} T_{+}^{F} 
\right)~d \sigma,\label{e45a}
\end{eqnarray}
where
\begin{equation}
T_{+}^{F}=\frac{i}{2} \left( \psi _{+}^{j, \mu}  \psi_{+j, \mu} -
\phi_{+}^{k, \mu}  \phi_{+ k, \mu} \right).
\end{equation}
The Majorana fermions are distinctly different with labels $j$ 
or $k$. The coefficient $k$ of the
total derivative of the last term in equation (\ref{e45a}) 
is $ \frac{1}{ \sqrt 6}$. 
$J=(1+ \frac{1}{ \sqrt 6})$. The normal ordering constant $a$ 
can be calculated from the values
for single bosonic and fermionic degrees of freedom~\cite{Green87} 
and in either case $a$=-1. 

\section{Virasoro generators and Physical states}
For a brief outline, let $L_m$, $G_r$ and $F_m$ be the Super 
Virasoro generators of energy, momenta and currents. 
Let $\alpha$'s denote the quanta of $X^{\mu}$ fields, b's and 
b$^{\prime}$'s denote quanta of $\psi$ and $\phi$ fields in 
NS formulation and d, d$^{\prime}$ in R formulation. Then,
\begin{eqnarray}
L_m&=&\frac{1}{\pi}\int_{-\pi}^{\pi}d\sigma e^{im\sigma}
T_{++}\nonumber\\
&= &\frac{1}{2}\sum^{\infty}_{-\infty}:
\alpha_{-n}\cdot\alpha_{m+n}: +\frac{1}{2}
\sum_{r\in z+\frac{1}{2}}(r+\frac{1}{2}m): 
(b_{-r} \cdot b_{m+r} - b_{-r}' \cdot b_{m+r}'):,~~~~~~~~~NS
\nonumber\\
&=&\frac{1}{2}\sum^{\infty}_{-\infty}:\alpha_{-n}
\cdot\alpha_{m+n}: +\frac{1}{2}
\sum^{\infty}_{n=-\infty}(n+\frac{1}{2}m): 
(d_{-n} \cdot d_{m+n} - d_{-n}'
\cdot d_{m+n}'):~, ~~~~~~~R\\
G_r &=&\frac{\sqrt{2}}{\pi}\int_{-\pi}^{\pi}d\sigma 
e^{ir\sigma}J_{+}=
\sum_{n=-\infty}^{\infty}\alpha_{-n}\cdot 
\left( e^jb_{n+r,j}- 
e^kb'_{n+r,k}\right ), ~~~~~~~~~~~~~~~~~~~~~~~~~~~~~~ NS\\
\text{and}\nonumber\\
F_m &=&\sum_{-\infty}^{\infty} \alpha_{-n}\cdot 
\left( e^jd_{n+m,j}- e^kd'_{n+m,k}\right ) .~~~~~~~~~ 
~~~~~~~~~~~~~~~~~~~~~~~~~~~~~~~~~~~~~~~~~~~~~~~~ R\label{e18}
\end{eqnarray}
and satisfy the super Virasoro algebra with central 
charge $C=26$ for the action of equation (\ref{e7}),
\begin{eqnarray}
\left [L_m , L_n\right ] 
& = &(m-n)L_{m+n} +\frac{C}{12}(m^3-m)\delta_{m,-n},\\
\left [L_m , G_r\right ] 
& = &(\frac{1}{2}m-r)G_{m+r}, 
~~~~~~~~~~~~~~~~~~~~~~~~~~~~~~~~~~~~~~~~~~~~~NS\\
\{G_r , G_s\} & =& 2L_{s+r} +\frac{C}{3}(r^2-
\frac{1}{4})\delta_{r,-s},\label{e19}\\
\left [L_m , F_n\right ] 
& = & (\frac{1}{2}m-n)F_{m+n},~~~~~~~~~~~~~~~~~~~~
~~~~~~~~~~~~~~~~~~~~~~~~~~R\\
\{F_m, F_n\} & = & 2L_{m+n} +\frac{C}{3}(m^2-1)
\delta_{m,-n},\;\;\;\; m\neq 0.\label{e19a}
\end{eqnarray}
Equations (\ref{e19}) and (\ref{e19a}) can be 
obtained using Jacobi identity.

This is also known that the normal ordering constant of
$L_o$ is equal to one and we define the physical 
states, satisfying
\begin{eqnarray}
(L_o-1)|\phi>&=&0,~~~L_m|\phi>=0,~~~G_r|\phi>
=0~~~ \text{for}~~~~ r,m>0,\;\;\;NS\;\;\;\text{Bosonic}\label{e20}\\
 L_m|\psi>&=&F_m|\psi>=0,\;\;\;\;\;\;\;\;\;\;\;\;\;\;\;
\text{for}\;\;\;\;\; m>0,\;\;\;\;\;\;\;\; ~~~~~~~~:R\;\;\;\;
\text{Fermionic}\label{e21}\\
and~~~(L_o-1)|\psi>_{\alpha}&=&(F_o^2-1)|\psi>_{\alpha}=0.\label{e22}
\end{eqnarray}
So we have,
\begin{equation}
(F_o +1)|\psi_+>_{\alpha}=0\;\;\;\; 
\text{and}\;\;\; (F_o-1)|\psi_->_{\alpha}=0.~~~~~~~~~~:R\label{e23}
\end{equation}
These conditions shall make the string model ghost free.
It can be seen in a very simple way. Applying $L_o$ condition,
the state  $\alpha_{-1}^{\mu}|0,k>$ is massless. 
The $L_1$ constraint gives the
Lorentz condition $k^{\mu}|0,k>=0$, implying a transverse 
photon and with
 $\alpha_{-1}^0|\phi>=0$ as Gupta-Bleuler condition. 
Applying $L_2,~~L_3~~ ....$, constraints,
one obtains $\alpha_{m}^0|\phi>=0$. Further, 
since $[\alpha_{-1}^0 , G_{r+1}]|\phi>=0$,~~
we have $b_{r,j}^0|\phi>=0$ and $b_{r,k}^{'0}|\phi>=0$. 
All the time components are 
eliminated from Fock space.

\section{BRST charge, tachyonlessness and modular invariance}
In order to prove the nilpotency of BSRT charge, we note that
the conformal dimension of $\gamma$ is `-1/2' and that 
of $\beta$ is `3/2', which can be deduced
from equation (\ref{e42a}). We can now proceed to write the 
nilpotent BRST charge. The part of the charge which comes 
from the usual conformal Lie algebra technique, is
\begin{equation}
(Q_1)^{NS,R} = \sum (L_{-m}c_m)^{NS,R} -\frac{1}{2}
\sum (m-n) : c_{-m}c_{-n}b_{m+n}:~ -
~a~ c_0 ;~~~~~Q^2_1=0 ~~~\text{for} ~a=1.
\end{equation}

Using the graded Lie algebra, we get the additional BRST 
charge, in a straight forward way, 
in NS and R,
\begin{eqnarray}
Q_{NS}'&=&\sum G_{-r}\gamma_r -\sum\gamma_{-r}\gamma_{-s}b_{r+s},\\
Q_{R}'&=&\sum F_{-m}\gamma_m -\sum\gamma_{-m}\gamma_{-n}b_{n+m},\\
and~~~~Q_{BRST}=Q_1 + Q',
\end{eqnarray}
is such that~~$Q_{BRST}^2=0$~~in both the NS and the 
R sector~\cite{Deo03}. In
proving $\{Q',Q'\} + 2\{Q_1,Q'\}=0$, we have used the Fourier
transforms, the wave equations and integration by parts such that
\begin{equation}
\sum\sum r^2\gamma_r\gamma_s\delta_{r,-s}
=\sum_r\sum_s\gamma_r\gamma_s\delta_{r,-s}=0.
\end{equation}
Thus the theory is unitary and ghost free.
There are no harmful effective tachyons in the model, eventhough 
\begin{eqnarray}
\alpha'M^2&=& -1, -\frac{1}{2}, 0 , \frac{1}{2}, 1, 
\frac{3}{2},..........NS\\
\text{and}\nonumber \\
\alpha'M^2&=& -1, 0 , 1, 2, 3,...................R.
\end{eqnarray}
The G.S.O. projection eliminates the half integral values. 
The tachyonic self energy of 
bosonic sector $<0|(L_o-1)^{-1}|0>$ is cancelled 
by $-<0|(F_o+1)^{-1}(F_o-1)^{-1}|0>_R$,
the negative sign being due to the fermionic loop. 
Such tadpole cancellations have been
noted also by Chattaraputi et al in reference~\cite{Chattaraputi021}.
One can proceed a step further and write down
the world sheet supersymmetric charge
\begin{equation}
Q=\frac{i}{\pi}\int_0^{\pi} \rho^0\rho^{\dag\alpha}
\partial_{\alpha}X^{\mu}\Psi_{\mu} d\sigma,
\end{equation}
and find, as it must,
\begin{equation}
 \sum\{Q_{\alpha}^{\dag}, Q_{\alpha}\} 
=2H ~~~~~\text{and} ~~~~~ \sum_{\alpha} |Q_{\alpha}
|\phi_o>|^2 = 2 <\phi_o|H|\phi_o> .
\end{equation}
The ground state is of zero energy. There are no 
overall tachyons in this Superstring.

To prove the modular invariance, one has to use the G.S.O. 
condition. In covariant
formulation, the number of degrees of freedom of fermions 
is the number obtained after substraction
of constraints from the total number. In our case, 
the total number is 44 and there are four
constraints. So the physical fermionic modes are 40. 
The partition function $Z$ can be found by
putting the 8 such fermions($2^3$ of $SO$(6)~) in each 
of the five boxes. This is a multiplication of 
spin structure of eight fermions $A_8$, five times,
\begin{equation}
A_8(\tau) = (\Theta_3 (\tau)/\eta(\tau))^{4} -
(\Theta_2 (\tau)/\eta(\tau))^{4}-(\Theta_4(\tau)/\eta(\tau))^{4},
\end{equation}
and
\begin{equation}
A_8(1+\tau)=-A(\tau)=A(-\frac{1}{\tau}),
\end{equation}
where the $\Theta(\tau)$'s are the Jacobi theta function 
and ~~$\eta(\tau)$ is
Dedekind eta function. Due to the Jacobi relation,
$A_8(\tau)=0$, the entire partition function, 
which is the product of all constituent partition 
functions~\cite{Deo103} of the model, vanishes.
Thus all the criteria prescribed in standard 
references are satisfied for this superstring. 

\section{Gauge Symmetry and the SUSY standard Model}

The zero mass particle spectrum of the quantised 
action(\ref{e7}) is very large. They are 
scalars, vectors and tensorial in nature. 
In reference~\cite{Deo04} with Maharana, we have 
shown that the massless excitations of the standard 
model can be found from this model. There 
are also the graviton and the gravitino~\cite{Deo104}. 
For the gauge symmetry, one has to find massless vector 
bosons which are generators of a group. We follow the 
work of Li~\cite{Li72}.

Consider $O(n)$. There are $\frac{1}{2}n(n-1)$ 
generators represented by 
\begin{equation}
L_{ij}= X_i\frac{\partial}{\partial X_j}-X_j
\frac{\partial}{\partial X_i}, ~~i,j=1,....,n.
\end{equation}
Using 
\[ \left [ \frac{\partial}{\partial X_i}, X_j\right ] 
=\delta_{ij}, \]
the Lie algebra, which is the commutator relation 
among the generators is
\begin{equation}
\left [ L_{ij}, L_{kl} \right ] 
= \delta_{jk} L_{il}+\delta_{il} L_{jk}-
\delta_{ik} L_{jl}-\delta_{jl} L_{ik}.
\end{equation}
Hence one must have $\frac{1}{2}n(n-1)$ vector gauge 
bosons $W^{\mu}_{ij}$ with the transformation law
\begin{equation}
W^{\mu}_{ij}\rightarrow W^{\mu}_{ij}+\epsilon_{ik} 
W^{\mu}_{kj}+ \epsilon_{jl}  W^{\mu}_{li}, ~~~~
W^{\mu}_{ij}= -W^{\mu}_{ji},
\end{equation}
where $\epsilon_{ij}= - \epsilon_{ji}$ are the infinitesimal 
parameters which characterise such rotation in
$O(n)$. Under gauge transformation of second kind,
\begin{equation}
W^{\mu}_{ij}\rightarrow W^{\mu}_{ij}+\epsilon_{ik} 
W^{\mu}_{kj}+ \epsilon_{jl}  W^{\mu}_{li}+\frac{1}{g}
\partial^{\mu}\epsilon_{ij}.
\end{equation}
The Yang-Mills Lagrangian is then written as
\begin{equation}
L=-\frac{1}{4}|F^{\mu\nu}_{ij}|^2,
\end{equation}
with 
\begin{equation}
F^{\mu\nu}_{ij} = \partial^{\mu}W^{\nu}_{ij}-
\partial^{\nu}W^{\mu}_{ij}+g\left (W^{\mu}_{ik}W^{\nu}_{kj}
-W^{\nu}_{ik}W^{\mu}_{kj}\right ),
\end{equation}
$F^{\mu\nu}_{ij}$ has the obvious properties, namely,
\begin{equation}
\Box F^{\mu\nu}_{ij}=0,~~~~ \partial_{\mu}
F^{\mu\nu}_{ij}=\partial_{\nu}F^{\mu\nu}_{ij}=0,
~~~~~~~ F^{\mu\mu}_{ij}=0.
\end{equation}

There are two sets of field strength tensors 
which are found in the model, one for $SO(6)$  and the other
for $SO(5)$. Since $\Box F^{\mu\nu}_{ij}=0 $, one can 
take plane wave solution and write 
\begin{equation}
F^{\mu\nu}_{ij}(x)=F^{\mu\nu}_{ij}(p)~e^{ipx},
\end{equation}
so that,
\begin{equation}
p^2 F^{\mu\nu}_{ij}(p)=p_{\mu}F^{\mu\nu}_{ij}(p)
=p_{\nu}F^{\mu\nu}_{ij}(p)=
F^{\nu\nu}_{ij}(p)=0~~~~~and~~~~~
F^{\mu\nu}_{ij}(p)=-F^{\mu\nu}_{ji}(p).
\end{equation}
These are physical state conditions (\ref{e20})-(\ref{e22}) as well.
\begin{equation}
L_0 F^{\mu\nu}_{ij}(p)=0,~~~~~~~~G_{\frac{1}{2}} F^{\mu\nu}_{ij}(p)=0,
\text{and}~~~~~~L_1 F^{\mu\nu}_{ij}(p)=0.\label{e71}
\end{equation}

The field strength tensor, satisfying (\ref{e71}), is found to be
\begin{equation}
F^{\mu\nu}_{ij}(p)= b^{\mu\dag}_i~b^{\nu\dag}_j |0,p> 
+ \epsilon_{ij}(p^{\mu}\alpha_{-1}^{\nu}-p^{\nu}
\alpha_{-1}^{\mu})|0,p>,\label{e71a}
\end{equation}
with $\epsilon_{ij} = e_i^{\alpha} e_j^{\beta} 
\varepsilon_{\alpha\beta},
~~(i,j)$=1,...,6~for O(6) and 1,...,5 for O(5) 
with $b'$ replaced by $b$.
For simplicity, we drop the $\dag$'s. 
In terms of the excitation quanta of the string, 
the vector generators are
\begin{equation}
W^{\mu}_{ij}=\frac{1}{\sqrt{2ng}}~n_{\kappa}
\epsilon^{\kappa\mu\nu\sigma}b_{\nu,i} b_{\sigma,j} 
+ \epsilon_{ij}\alpha_{-1}^{\mu},
\end{equation}
 where $n_{\kappa}$ is the time-like four vector 
and can be taken as (1,0,0,0). One finds that
\begin{eqnarray}
\partial^{\mu}W^{\nu}_{ij}-\partial^{\nu}W^{\mu}_{ij}
&=&p^{\mu}W^{\nu}_{ij}-p^{\nu}W^{\mu}_{ij}\nonumber\\
&=&\frac{1}{\sqrt{2ng}}\left ( n_{\kappa}
\epsilon^{\kappa\nu\lambda\sigma}p^{\mu}
-n_{\kappa}\epsilon^{\kappa\mu\lambda\sigma}
p^{\nu}\right )b_{\lambda,i} b_{\sigma,j} 
+ \epsilon_{ij}(p^{\mu}\alpha_{-1}^{\nu} 
- p^{\nu}\alpha^{\mu}_{-1}).
\end{eqnarray}
As $\mu$ must be equal to $\nu$, if $\kappa,~\lambda ~
\text{and}~ \sigma$ are the same,
the first term vanishes and 
\begin{eqnarray}
g\left (W^{\mu}_{ik}W^{\nu}_{kj} 
- W^{\nu}_{ik}W^{\mu}_{kj}\right )
&=&\frac{1}{2n}\left (  n_{\kappa}~
\epsilon^{\kappa\mu\lambda\sigma}~
b_{\lambda,i} b_{\sigma,k}
~n_{\kappa'}~\epsilon^{\kappa'\nu\lambda'\sigma'}~
b_{\lambda',k} b_{\sigma,j}
-~~~\mu\leftrightarrow \nu\right )
+ \epsilon_{ik}\epsilon_{kj}\left [\alpha_{-1}^{\mu}, 
\alpha_{-1}^{\nu}\right ]\nonumber\\
&=&b_i^{\mu}b_j^{\nu}.
\end{eqnarray}
We have used 
\[
\left \{ b_{\lambda'k}, b_{\sigma k}\right \} =
\eta_{\lambda'\sigma}\delta_{kk}=~n~\eta_{\lambda'\sigma},
\]
and the creation operators $\alpha_{-1}^{\mu}, \alpha_{-1}^{\nu}$ commute.
Equation (\ref{e71a}) is referred as the field strength.
Since the product of pairs of $b$ and $b'$ commute, the gauge group of 
the action~(\ref{e7}) is the product group $SO(6)\otimes SO(5)$. 
This is same as the symmetry group of the action (\ref{e7}).

To descend to the standard model group 
$SU_C(3)\otimes SU_L(2)\otimes U_Y(1)$, one normaly 
introduced by Higgs which break gauge symmetry and supersymmetry. 
However, if one uses the method of symmetry breaking 
by using Wilson's loops, supersymmetry remains 
intact but the gauge symmetry is broken. The Wilson loop is
\begin{equation}
U_{\gamma}= P~exp\left (\oint_{\gamma} A_{\mu}\;dx^{\mu}\right ).
\end{equation}
$P$ represents the ordering of each term with respect 
to the closed path $\gamma$. SO(6)=SU(4)
descends to $SU_C(3)\otimes U_{B-L}(1)$. This breaking can be 
accomplished by choosing one element of $U_0$ of SU(4), such that
\begin{equation}
U_0^2=1.
\end{equation}
The element generates the permutation group $Z_2$. Thus
\begin{equation}
\frac{SO(6)}{Z_2}= SU_C(3)\otimes U_{B-L}(1),
\end{equation}
without breaking supersymmetry. Similarly,
$SO(5)\rightarrow SO(3)\otimes SO(2)=SU(2)\otimes U(1)$. 
We have,
\begin{equation}
\frac{SO(5)}{Z_2}=SU(2)\otimes U(1).\label{e80a}
\end{equation}
Thus
\begin{equation}
\frac{SO(6) \otimes SO(5)}{Z_2 \otimes Z_2}=
SU_C(3)\otimes U_{B-L}(1)\otimes U_R(1) \otimes SU_L(2),
\end{equation}
making an identification with the usual low energy phenomenology. 
But this is not the  standard model. We have an additional U(1). 
However, there is an instance in $E_6$, where there 
is a reduction of rank by one and several U(1)'s. 
Following the same idea~\cite{Green87}, 
we may take
\begin{equation}
U_{\gamma}=(\alpha_{\gamma}) \otimes \left(\begin{array}{ccc}
\beta_{\gamma}&&\\&\beta_{\gamma}&\\&&
\beta^{-2}_{\gamma}\end{array}\right) \otimes 
\left( \begin{array}{ccc}\delta_{\gamma} &&\\&
\delta^{-1}_{\gamma}&\\&&
\end{array}\right).
\end{equation} 
$\alpha_{\gamma}^3$ =1 such that $\alpha_{\gamma}$ is the cube 
root of unity. This structure lowers the rank by one. We have,
\begin{equation}
\frac{SO(6)\otimes SO(5)}{Z_3}=SU_C(3)\otimes SU_L(2)\otimes  U_Y(1),
\end{equation}
and find the supersymmetric standard model.

We now elaborately discuss  the $Z_3$, described by ~\cite{Mishra92},
\begin{equation}
g(\theta_1, \theta_2,\theta_3) =
\left(\frac{2\pi}{3}-2\theta_1,\frac{2\pi}{3}+\theta_2, 
\frac{2\pi}{3}+\theta_3 \right).
\end{equation}
For the first Wilson loop, the angle integral for 
$\theta_1 =\frac{2\pi}{9}$ to $\frac{2\pi}
{3} - \frac{4\pi}{9} = \frac{2\pi}{9}$, so that the first loop 
integral vanishes. $\theta_2 =0$ to
$2\pi -\frac{2\pi}{3} =\frac{4\pi}{3}$ for the second loop, is described 
by a length parameter $R$,
$\theta_3 =0$ to $2\pi -\frac{2\pi}{3}= \frac{4\pi}{3}$ 
for the remaining loop with the same $R$. We take the polar 
components of the gauge 
fields as non-zero constants, as given below.
\[ g A_{\theta_2}^{15} = \vartheta_{15}\] 
for SO(6) = SU(4) 
for which the diagonal generator $t_{15}$ breaks the symmetry and
\[ g'A_{\theta_3}^{10} =\vartheta'_{10} \] 
for SO(5), the diagonal generator being $t'_{10}$. 
The generators of both SO(6) 
and SO(5)  are $4\times 4$ matrices. We can write the $Z_3$ group as
\begin{equation}
T = T_{\theta_1}T_{\theta_2}T_{\theta_3}.
\end{equation}
We have $T_{\theta_1}$ =1.
This leaves the unbroken symmetry $SU(3)\times SU(2)$ untouched
\begin{equation}
T_{\theta_2}= exp\left( i~t_{15}\int_o^{\frac{4\pi}{3}}\vartheta_{15} R 
d\theta_2\right ).
\end{equation}
$T_{\theta_2} \neq 1$ breaks the SU(4) symmetry. Again,
\begin{equation}
T_{\theta_3}= exp\left( i~t'_{10}\int_o^{\frac{4\pi}{3}}
\vartheta'_{10} R d\theta_3 \right ).
\end{equation}
$T_{\theta_3} \neq 1$ breaks the SO(5) symmetry.
But the remaining product of $Z_3$ is
\begin{equation}
T_{\theta_2} T_{\theta_3} =exp \left(
  i\int_o^{\frac{4\pi}{3}}(\vartheta'_{10}t'_{10} +
\vartheta_{15}t_{15}) R d\theta \right ).
\end{equation}
$\vartheta_{15}$ and $\vartheta'_{10}$~~ are 
arbitrary constants. We can choose them in such a way that ~~
\[t_{15}\vartheta_{15} + t'_{10}\vartheta'_{10}=0, 
\frac{3}{2R}, ... ..\] 
The term in the exponential is zero or multiples of $2\pi i$. 
Thus $T= U(1)$ and equation (\ref{e80a})
is obtained, reducing the rank by one.

\section{Replication of families and concluding remarks}
The residual supersymmetry at electroweak scale is 
$Z_3\times SU_C(3)\times SU_L(2)\times U_Y(1)$
and will be denoted by GZ. To find the number of generations 
of the theory $n_G$, we have to calculate Euler number $\chi$, since
\begin{equation}
n_G=\frac{1}{2}\chi(GZ).\label{e100}
\end{equation}
The following formulas are relevant. The space with known
Euler number is
\begin{equation}
CP_N=\frac{SU(N+1)}{U(N)}\label{e101}
\end{equation}
with
\begin{equation}
\chi(CP_N)=(N+1).\label{e102}
\end{equation}
Using the the above equations (\ref{e101}) and (\ref{e102}), 
we find $\chi(GZ)$ = 6 and $n_G$=3. These group structure gives 
the number of generations to be 3. This number is topological because
of Dirac index theorem. The Dirac equation, in general, 
can have zero modes,
\begin{equation}
\gamma\cdot p~\psi=p\!\!\!/~\psi =D\!\!\!\!/~ \psi =0.\label{e103}
\end{equation}
The index of this operator is equal to the difference between the 
positive, $n_+$ and negative, $n_-$ chiralities of the zero modes. 
The exact theorem, on its relation with $n_G$, is 
\begin{equation}
n_G=\frac{1}{2}\chi= index( D\!\!\!\!/~)=n_+-n_-,\label{e104}
\end{equation}
of the zero modes. It is necessary to find the 
massless four dimensional Dirac spinors. The
creation operators of the Ramond sector is
\begin{equation}
D^{\mu\dagger}=e^jd_{-1,j} -e^kd_{-1,k},\label{e105}
\end{equation}
is such that a zeromass spinorial state isi~\cite{Deo104}
\begin{equation}
\phi_{o+}=D_{-1}^{\mu}|0>u_{\mu}=D^{\mu\dagger}|0>u_{\mu}\label{e106}
,\end{equation}
$u_{\mu}$ is a spinor four vector. Similarly, there is another state
\begin{equation}
\phi_{o-}=\alpha_{-1}^{\mu}|0>v_{\mu}=\alpha^{\mu\dagger}|0>
v_{\mu}\label{e107}
,\end{equation}
for the coordinate excitation of the Ramond sector, $v_{\mu}$ 
is also another
four vector spinor. Both ($u_{\mu},v_{\mu}$) are in four 
dimensions. They can be distinguished by (\ref{e8})
\begin{eqnarray}
\gamma_5 u_{\mu}=u_{\mu},\label{e109}\\
and~~~~~\gamma_5 v_{\mu}=-v_{\mu}.\label{e110}
\end{eqnarray}
Together, they can also be considered as a four component spin vector
$\psi_{\mu}$. Since $F_o$ is essentially the Dirac gamma matrix ($\gamma$),
the condition $F_o\psi_{\mu}$ gives the zeromass spinor equation
\begin{equation}
\gamma\cdot p~\psi_{\mu}=0.\label{e111}
\end{equation}
 The states $\phi_{0,\pm}$ contains not only a spin-$\frac{1}{2}$ but 
also a spin-$\frac{3}{2}$ state. But one can covariantly separate 
out the Dirac spin-$\frac{1}{2}$ equation as given in~\cite{Gliozzi76}.
If $\psi_{Dirac}$ is any Dirac spinor in 4-dimensions satisfying
$p\!\!\!/\psi$=0, then the spin-$\frac{1}{2}$ component is
\begin{equation}
\psi_{\mu}^{\frac{1}{2}}
=\frac{1}{2}\left [ \gamma_{\mu}-p_{\mu}
(\gamma\cdot \bar{p})\right ]\psi_{Dirac},\label{e112}
\end{equation}
where the momentum $\bar{p}_{\mu}$ is conjugate to $p_{\mu}$
with $p^2=\bar{p}^2=0$ and $p\cdot\bar{p}$=1.

The zero mass modes of Dirac objects in standard model
are grouped into three families(generations). They are
\begin{eqnarray}
\left (
\begin{array}{c}
u\\
d\\
\nu_e\\ 
e\\
\end{array}
\right ),
\left (
\begin{array}{c}
c\\
s\\
\nu_{\mu}\\
\mu\\
\end{array}
\right ),~~~and~~~~
\left (
\begin{array}{c}
t\\
b\\
\nu_{\tau}\\
\tau\\
\end{array}
\right )
\end{eqnarray}
The left handed ones are doublets and right handed 
ones are singlets. The quarks are colored. There are 
several fermionic zero modes, 24 are with +ve helicity and 21
with negative helicity.
Thus $n_+-n_-$=3 as per topological
findings. So there have to be only three fermions(neutrinos
with unpaired helicity) in the standard model.

Thus, we have made the successful attempt 
in constructing an anomaly free $N=1, D=4$ superstring 
from bosonic string with the gauge symmetry 
$SO(6)\otimes SO(5)$ which, with the help of Wilson 
loops, descend to the SUSY standard model. There 
are just three generations. This
is a very important result which can help not 
only in  phenomenology but also
in finding the correct dynamics of interacting string 
following up the  bosonic equivalence.

It has been summarised by Mohapatra~\cite{Mohapatra96}
that the uniqueness of the three generations model
constructed by Tian and Yau, is the triumph of Calabi-Yau
compactification procedure and  all such models are equivalent.
We propose to study if the present model also falls in the same category.

\begin{acknowledgments}
I have been profited from a discussion with Prof L. Maharana 
which is thankfully acknowledged.
\end{acknowledgments}

\end{document}